\documentclass[prl,twocolumn,showpacs,amsmath,amssymb]{revtex4}

\usepackage{graphicx,dcolumn,bm}

\begin{document}

\title{Fluctuations and topological transitions of quantum Hall stripes: \\
nematics as anisotropic hexatics}

\author{A.M. Ettouhami}
\email{mouneim@physics.utoronto.ca}
\affiliation{Department of Physics, University of Toronto, 60 St. George St., Toronto M5S 1A7, 
Ontario, Canada}

\author{C.B. Doiron}
\affiliation{Department of Physics and Astronomy, University of Basel,
Klingelbergstrasse 82, CH-4056 Basel, Switzerland}

\author{R. C\^ot\'e} 
\affiliation{D\'epartement de Physique, Universit\'e de Sherbrooke, Qu\'ebec, Canada, J1K 2R1}

\date{\today}

\begin{abstract}

We study fluctuations and topological melting transitions of 
quantum Hall stripes near half-filling of intermediate Landau levels. 
Taking the stripe state to be an anisotropic Wigner crystal (AWC)
allows us to identify the quantum Hall nematic state conjectured
in previous studies of the 2D electron gas
as an {\em anisotropic hexatic}. The transition 
temperature from the AWC to the quantum Hall nematic state is explicitly
calculated, and a tentative phase diagram for the 2D electron gas near half-filling
is suggested.

\end{abstract}



\maketitle

\begin{figure}[t]
\includegraphics[width=8cm, height=4.5cm]{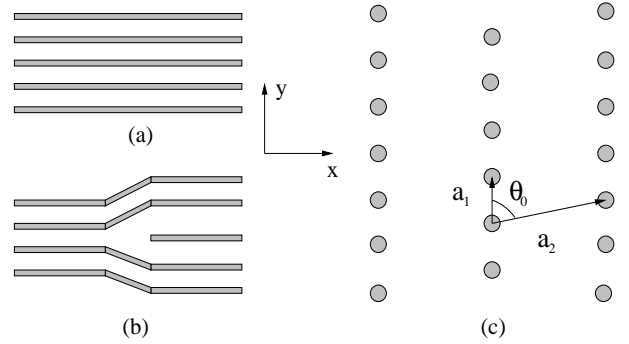}
\caption[]{Schematic view of (a) a smectic; (b) a smectic with an edge dislocation;
and (c) an anisotropic Wigner crystal where the guiding centers of the electrons 
occupy the sites of an anisotropic triangular lattice. 
}\label{Fig:lattice}
\end{figure}

{\em Introduction} -- Following theoretical predictions by Koulakov {\em et al.} \cite{Koulakov1996} 
that the ground state of the 2D electron gas near half filling of 
intermediate Landau levels (LLs), with index $N\ge 2$,
is a striped state, and the subsequent experimental observation by Lilly {\em et al.}\cite{Lilly1999} 
of strongly anisotropic dc resistivities in the above mentioned range of fillings, 
it has been suggested \cite{Kivelson1998,Fradkin1999} that the striped ground state of a two-dimensional 
electron gas (2DEG) at low temperature may be viewed as a ``quantum Hall smectic" (QHS), 
consisting of a weakly coupled stack of one-dimensional Luttinger liquids. This is a state that 
would only be stable at zero temperature, and which, by analogy with conventional liquid 
crystals \cite{Toner1981}, would give way through the proliferation of dislocations 
(see panels (a) and (b) of Fig. \ref{Fig:lattice}) to a ``nematic" state at nonzero temperatures 
\cite{Radzihovsky2002}, in which translational symmetry is restored but rotational symmetry 
is still broken. This electronic ``nematic" would then undergo a disclination unbinding transition 
into a fully isotropic fluid as temperature is raised above a critical temperature which has been 
estimated \cite{Fradkin2000,Wexler2001} following standard Kosterlitz-Thouless (KT) arguments 
\cite{Toner1981}.

In this paper, we want to examine an alternative picture, in which
the ground state of the 2DEG near half filling of intermediate LLs
is taken to be an anisotropic Wigner crystal (AWC), 
as suggested by Hartree-Fock (HF) \cite{Cote2000,Ettouhami2006}  
and renormalization group (RG) \cite{MacDonald2000} calculations.
In this case, we find that dislocations melt the AWC at a nonzero
temperature (that we shall explicitly evaluate below) 
into a ``nematic" state with quasi-long-range orientational correlations,
which we argue is nothing more than an {\em anisotropic hexatic}. 
Our results for the melting temperature of the AWC are consistent
with experiments and with the idea of quantum Hall ``nematics" 
conjectured in Refs. \onlinecite{Kivelson1998,Fradkin1999}.

{\em Fluctuations of quantum Hall Wigner crystals} --
In what follows, we shall be interested in the elastic fluctuations of quantum Hall
Wigner crystals. To fix ideas, we shall focus on the AWC which was found to 
minimize the cohesive energy of the 2DEG near half filling
if Ref. \onlinecite{Ettouhami2006},
and which is described by the lattice vectors ${\bf R}_{n_1 n_2}= n_1{\bf a}_1 + n_2{\bf a}_2$, 
where ${\bf a}_1 = 2\alpha\hat{\bf y}$ and ${\bf a}_2 = \alpha\hat{\bf y} + \beta\hat{\bf x}$,
and where $\alpha=a\sqrt{1-\varepsilon}/2$ and $\beta=\sqrt{3}a/2\sqrt{1-\varepsilon}$ ($n_1$ and $n_2$ 
being integers). In these expressions, $\varepsilon$ is a positive parameter such that 
$0\leq \varepsilon < 1$ which quantifies the degree of anisotropy of the lattice at a given partial 
filling factor $\nu^*$, and which was determined through minimization of
the cohesive energy of the system in Ref. \onlinecite{Ettouhami2006}; and $a=\ell(4\pi/\sqrt{3}\nu^*)$
is the average spacing of a hexagonal lattice with $\varepsilon=0$ at the same value of $\nu^*$.
The elastic properties of such an anisotropic crystal can be described by an elastic Hamiltonian
of the form ($\alpha,\beta = x,\,y$): 
\begin{equation}
H = \frac{1}{2}\int d{\bf r}d{\bf r}'\, C_{\alpha\beta\gamma\delta}({\bf r}-{\bf r}')
\,u_{\alpha\beta}({\bf r}) u_{\gamma\delta}({\bf r}'),
\end{equation}
where $u_{\alpha\beta}({\bf r})=\frac{1}{2}(\partial_\alpha u_\beta 
+ \partial_\beta u_\alpha)$ is the linear strain tensor,
(${\bf u}({\bf r})$ being the displacement field).
For the particular case of a two-dimensional AWC, there are three compression moduli,
$c_{11}\equiv C_{1111}$, $c_{22}\equiv C_{2222}$ and $c_{12}\equiv C_{1122}$,
and a single shear modulus $c_{66}\equiv C_{1212}$.

The elastic fluctuations of the above AWC, taking into account the Lorentz-force
dynamics imposed by the external magnetic field, can be described by the Gaussian action:
\begin{eqnarray}
S = \frac{1}{2}\sum_{n=-\infty}^\infty\int_{\bf q} u_\alpha({\bf q},\omega_n) 
D_{\alpha\beta}({\bf q},\omega_n)
u_\beta(-{\bf q},-\omega_n),
\label{Eq:Action}
\end{eqnarray}
where the dynamical matrix $D_{\alpha\beta}({\bf q},\omega_n)$ is given by:
\begin{equation}
D_{\alpha\beta}({\bf q},\omega_n) = \Phi_{\alpha\beta}({\bf q}) 
+ \rho_m\omega_n^2\delta_{\alpha\beta} + \varepsilon_{\alpha\beta}\rho_m \omega_c\omega_n.
\end{equation}
In the above expression, $\omega_n=2\pi n k_BT/\hbar$ is a Matsubara frequency
($k_B$ being Boltzmann's constant and $T$ being temperature),
$\rho_m$ is a mass density, $\omega_c$ is the cyclotron 
frequency, $\varepsilon_{\alpha\beta}$ is the two-dimensional version of the antisymmetric 
Levi-Civita tensor, and we have introduced the elastic matrix $\Phi_{\alpha\beta}({\bf q})$, 
which has the following matrix elements:
\begin{subequations}
\begin{align}
& \Phi_{xx}({\bf q}) = c_{11}q_x^2 + c_{66}q_y^2,
\\
& \Phi_{yy}({\bf q}) = c_{22}q_y^2 + c_{66}q_x^2,
\\
& \Phi_{xy}({\bf q}) = \Phi_{yx}({\bf q}) = \big(c_{12} + c_{66}\big)q_xq_y.
\end{align}
\end{subequations}

From Eq. (\ref{Eq:Action}), we can easily derive the following expression for the
two-point correlation function $\langle u_\alpha({\bf q},\omega_n)u_\beta({\bf q}',\omega_l)\rangle$:
\begin{equation}
\langle u_\alpha({\bf q},\omega_n)u_\beta({\bf q}',\omega_l)\rangle 
= (2\pi)^2\delta_{n,-l}\delta({\bf q}+{\bf q}')
\hbar G_{\alpha\beta}({\bf q},\omega_n),
\end{equation}
where $\delta_{n,l}$ is the Kronecker symbol, and
where the propagator $G_{\alpha\beta}$ has the following elements (we sum over repeated indices):
\begin{equation}
G_{\alpha\beta}({\bf q},\omega_n)  =  \frac{\varepsilon_{\alpha\gamma}\varepsilon_{\beta\delta}
\Phi_{\gamma\delta}({\bf q}) 
+ \rho_m\omega_n^2\delta_{\alpha\beta} - \varepsilon_{\alpha\beta}\rho_m\omega_c\omega_n}{\mbox{Det}(D)}\,,
\end{equation}
with
\begin{align}
\mbox{Det}(D) & =  \rho_m^2\omega_n^4 + \rho_m \omega_n^2 \big[\rho_m\omega_c^2 + \Phi_{xx}({\bf q})
+ \Phi_{yy}({\bf q})\big] 
\nonumber\\
& + \Phi_{xx}({\bf q})\Phi_{yy}({\bf q})-\Phi_{xy}^2({\bf q}).
\end{align}
In real space, the mean squared displacement
$\langle u_\alpha({\bf r},\tau)u_\beta({\bf r},\tau)\rangle$
can be written in the form
$\langle u_\alpha({\bf r},\tau)u_\beta({\bf r},\tau)\rangle = k_BT 
\int_{\bf q} \tilde{G}_{\alpha\beta}({\bf q})$,
with the effective propagator $\tilde{G}_{\alpha\beta}({\bf q}) = \sum_{n=-\infty}^\infty 
G_{\alpha\beta}({\bf q},\omega_n)$.
Performing the Matsubara sum, we find that the static fluctuations of the quantum system at finite
temperatures can be described using the partition function
$Z_{cl} = \int [du({\bf r})]\,e^{-\tilde{H}/k_BT}$,
with the effective {\em classical} Hamiltonian:
\begin{equation}
\tilde{H} = \frac{1}{2}\int_{\bf q} u_\alpha({\bf q})
\big[\tilde{G}^{-1}({\bf q})\big]_{\alpha\beta}
u_\beta(-{\bf q}). 
\end{equation}
Knowledge of the effective propagator $\tilde{G}({\bf q})$
allows us to study the effect of Lorentz-force dynamics 
on the elastic properties (and hence on
possible topological transitions) of the system.
Calculating the inverse effective propagator $\tilde{G}^{-1}({\bf q})$ and expanding 
the resulting expression near $q=0$, we find that, up to terms of order $O(q^2)$, 
the form of the elastic propagator is identical to its zero field expression. 
This has the important consequence that the long wavelength 
elastic properties of the AWC will be qualitatively the same as in the absence of magnetic field.
We therefore expect the topological melting of the AWC 
to proceed in a standard (two-stage) way \cite{Ostlund1981}, 
as we now are going to describe.

\begin{figure}[t]
\includegraphics[width=8.09cm, height=5cm]{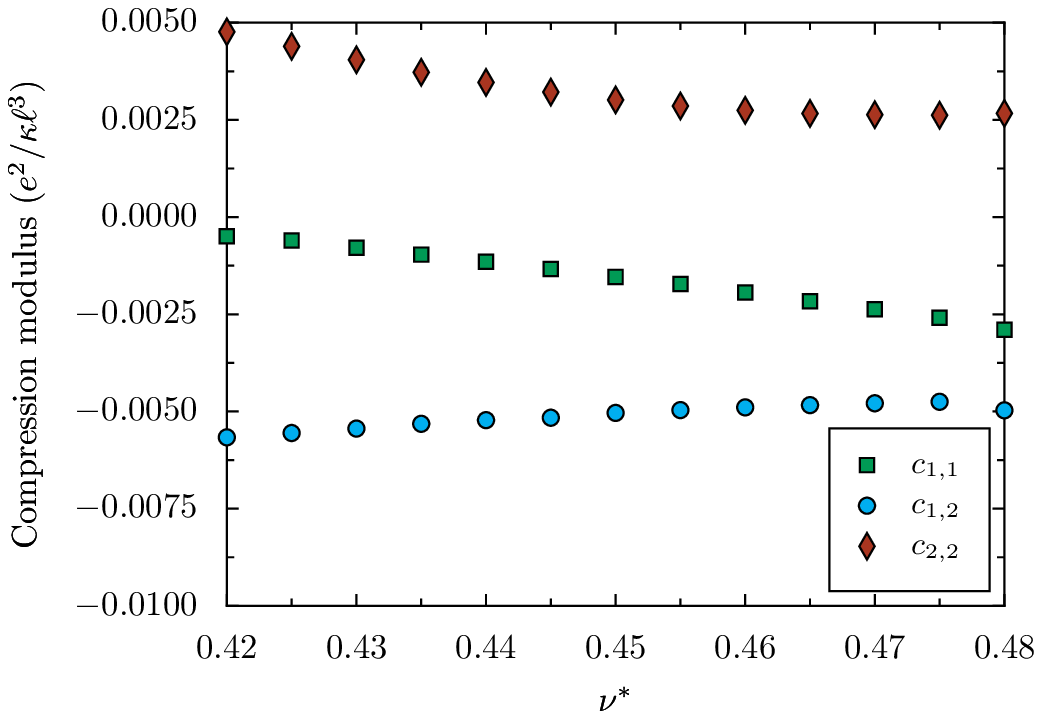}
\caption[]{(Color online) 
Finite parts $\tilde{c}_{ij}$, $i,\,j=1,\,2$, of the compression moduli for the anisotropic 
Wigner crystal as a function of the partial filling factor $\nu^*$ in Landau level $N=2$.
}\label{Fig:moduli}
\end{figure}

{\em Topological melting of anisotropic Wigner crystals} -- 
A major difference between isotropic and anisotropic Wigner crystals in two dimensions is that,
while in the former all six elementary dislocations differing by
the orientation of their Burger's vectors are equivalent, in the latter
two elementary equivalent dislocations (labeled type I)
have their Burger's vectors along a reflection symmetry axis
(i.e. along $\pm {\bf a}_1$ in Fig. 1), 
while four dislocations, equivalent to each other but inequivalent to the first type, lie at angles of 
$\pm\theta_0$ from the reflexion axis ($\theta_0$ here is the angle between ${\bf a}_1$ and ${\bf a}_2$, 
see Fig. \ref{Fig:lattice}). At any nonzero temperature, the solid phase has a finite density of 
tightly bound dislocation pairs. As the temperature is raised past a critical temperature $T_{c1}$, 
the pairs unbind and destroy the crystalline order. Since the two types I and II of dislocations 
are unequivalent, the defect mediated melting (DMM) process
will be governed by the type which has a lower nucleation energy. 
We therefore shall need to determine the elastic constants of the AWC
in order to find the energies of the two types of dislocations, so as to determine which
dislocation type unbinds first. 

\begin{figure}[t]
\includegraphics[width=8.09cm, height=5cm]{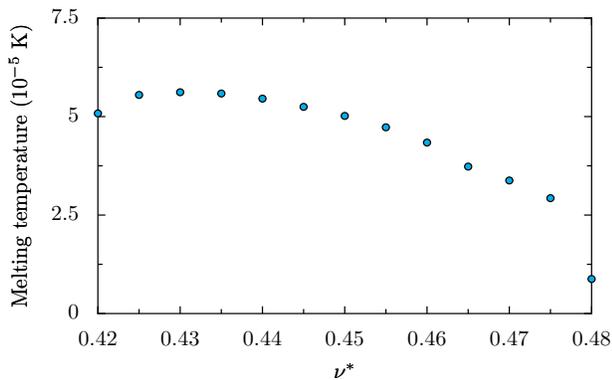}
\caption[]{(Color online) 
Critical temperature $T_{c1}$ for the dislocation mediated melting of the anisotropic 
Wigner crystal into a nematic-like state near half filling of Landau level $N=2$. 
}\label{Fig:Tm}
\end{figure}

For the AWC which is the object of study in this paper, 
the compression moduli $c_{ij}(q)$, $i,j=1,2$, can be written in the form
$c_{ij}({\bf q})  =  c(q) + \tilde{c}_{ij}$, ($i,j=1,2$),
where we separated out the leading (plasmonic) contribution \cite{Ettouhami2006prb} 
$c(q) = \left({e^2}/2\pi\kappa\ell^3\right)/{q\ell}$,
$e$ being the electronic charge, $\ell$ the magnetic length, and $\kappa$ the
dielectric constant of the host medium.
For a one-dimensional compression of the form ${\bf u}_1=u_0x\hat{\bf x}$, with
the number of electrons $N_e$ in Landau level $N$ kept fixed, if we denote by
$\nu_1^*=\nu^*/(1+u_0)$ the partial filling factor {\em of the compressed crystal}, 
it can be shown \cite{unpublished} that the constant part $\tilde{c}_{11}(\nu^*)$ 
is given by:
\begin{equation}
\tilde{c}_{11}\left(\nu^*\right) = \frac{(e \nu^*)^{2}}{2\pi\kappa\ell^{3}}
\left[ \nu^* \frac{d^{2}\mathcal{G}(\nu^*) }{d{\nu^*}^2}+2\frac{d\mathcal{G}\left(\nu^*\right)}{d\nu^*}
\right]_{\nu^* = \nu_1^*}.
\end{equation}
In the above expression, $\mathcal{G}(\nu)$ is the HF cohesive energy per electron
(in units of $e^2/\kappa\ell$), and is given by:
\begin{align}
\mathcal{G}(\nu^*) & = \frac{1}{2\nu^*}\sum_{\bf Q}\Big[ H_{N}\left( \mathbf{Q}\right) 
\left(1-\delta _{\mathbf{Q},0}\right)
- X_{N}\left( \mathbf{Q}\right) \Big] 
\left\vert \left\langle \rho\left( \mathbf{Q}\right) \right\rangle \right\vert ^{2},
\end{align}
where ${\bf Q}$ is a reciprocal lattice vector, and
$\rho\left({\bf Q}\right)$ is the guiding-center density operator,
which is determined self-consistently using the approach of Ref. \onlinecite{Cote2000},
and is related to the real density operator $n(\mathbf{Q})$ through the equation
(here $N_{\varphi}$ is the Landau level degeneracy and $L_{N}^{0}\left( x\right) $ is a generalized
Laguerre polynomial):
\begin{equation}
n(\mathbf{Q})=N_{\varphi }e^{-Q^{2}\ell ^{2}/4}
L_{N}^{0}\left(\frac{Q^{2}\ell ^{2}}{2}\right) \rho (\mathbf{Q}).
\end{equation}
On the other hand, the Hartree and Fock interactions are given by\cite{Cote2000}
($J_{0}\left( x\right) $ is the Bessel function of order zero): 
\begin{subequations}
\begin{align}
& H_{N}\left( \mathbf{q}\right) =\frac{1}{q\ell }e^{-q^{2}\ell^{2}/2}
\left[ L_{N}^{0}\left( \frac{q^{2}\ell ^{2}}{2}\right) \right] ^{2}, 
\\
& X_{N}\left( \mathbf{q}\right) =\sqrt{2}\int_{0}^{\infty }dx\,e^{-x^{2}}
\left[ L_{N}^{0}\left( x^{2}\right) \right] ^{2}J_{0}\left( \sqrt{2}x q\ell
\right).
\end{align}
\end{subequations}

The finite contributions $\tilde{c}_{22}$ and $\tilde{c}_{12}$
to the compression moduli ${c}_{22}$ and ${c}_{12}$ can be obtained
in a similar way by considering 1D and 2D uniform compressions of the form
${\bf u}=u_0y\hat{\bf y}$ and ${\bf u}=u_0(x\hat{\bf x} + y\hat{\bf y})$, respectively.
The results of these procedures, the details of which will be published elsewhere \cite{unpublished}, are 
shown in Fig. \ref{Fig:moduli}, where we plot the constant parts $\tilde{c}_{ij}$ ($i,\,j=1,\,2$)
of the compression moduli of the stripe crystal near half filling of LL $N=2$.

Let us now introduce the compliance tensor $S_{ijkl}$ such that 
$S_{ijkl}C_{klmn}=\frac{1}{2}(\delta_{im}\delta_{jn}
+ \delta_{in}\delta_{jm})$, with the four independent elements $s_{11}({\bf q})\equiv S_{1111}$,
$s_{22}({\bf q})\equiv S_{2222}$, $s_{12}({\bf q}) = s_{21}({\bf q}) \equiv S_{1122}$, 
and $s_{66}\equiv S_{1212}$. In the long wavelength limit, these take the values: 
\begin{align}
& s_{11} = s_{22} = -s_{12} = \frac{1}{ \tilde{c}_{11}+\tilde{c}_{22}-2\tilde{c}_{12} },
\quad s_{66}=\frac{1}{4c_{66}}. 
\end{align}
In terms of these compliances, we find that
the leading contribution to the energy of a dislocation of type $\alpha$ ($\alpha=I$ or $II$) is given by
\begin{equation}
E_\alpha = \frac{1}{2}K_\alpha \ln(L/a),
\end{equation}
with \cite{Ostlund1981}:
\begin{subequations}
\begin{align}
& K_I = a^2(1-\varepsilon)\sqrt{\frac{s_{22}}{s_{11}}}K,
\\
& K_{II}=\frac{1}{4}K_I\Big[1+\frac{3}{(1-\varepsilon)^2}\sqrt{\frac{s_{11}}{s_{22}}}\Big],
\end{align}
\end{subequations}
where we defined
\begin{equation}
K= \frac{1}{2\pi}[2s_{22}(\sqrt{s_{11}s_{22}} + s_{12} + 2s_{66})]^{1/2}.
\end{equation}
For the problem at hand, $s_{11}=s_{22}$ in the long wavelength limit, and hence we see that the ratio
$K_{II}/K_{I}$ is always larger than unity for $0<\varepsilon<1$. We thus see
that dislocations of type I are energetically less costly than type II dislocations, and will unbind
at the melting temperature $T_{c1}$ such that \cite{Ostlund1981}
$K_Ia^2/k_BT_{c1}=4$ (the value $4$ being universal), from which we obtain the melting temperature 
\begin{equation}
k_BT_{c1}=\frac{1}{4}\,a^2(1-\varepsilon)K_{11}.
\end{equation}
The resulting melting line of the AWC is plotted in Fig. \ref{Fig:Tm}.

\begin{figure}[t]
\includegraphics[width=8.09cm, height=5cm]{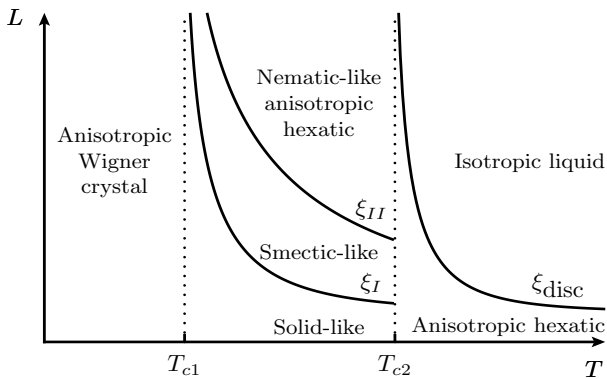}
\caption[]{ 
Tentative phase diagram showing the various relevant regimes for the melting of quantum Hall stripes
as a function of temperature $T$ and length scale $L$. $\xi_{disc}$
is the average spacing between disclinations. Below this
length scale above $T_{c2}$ the system retains the properties of an anisotropic hexatic.
Note that $T_{c1}$ here can be extremely small, and may be driven to zero by quantum fluctuations.
}\label{Fig:phasediagram}
\end{figure}

For temperatures higher than $T_{c1}$, the presence of type I dislocations screens the logarithmic 
interaction between type II dislocations, such that both dislocation types are free 
at long length scales\cite{Ostlund1981}. We expect, however, 
that the average separation $\xi_{II}$ of type II dislocations will be much larger
than the average separation $\xi_I$ between type I dislocations, 
with \cite{Ostlund1981}:
\begin{equation}
\xi_I\sim \exp(t^{-1/2}), \quad \xi_{II}\sim \xi_I^{p+1},
\end{equation} 
where $t\propto (T-T_{c1})$ and $p$ is a nonuniversal number between 0 and 2. 
Given that type I dislocations destroy translational order mainly along the direction of the stripes, 
we see that we can distinguish between three different regimes (see Fig. \ref{Fig:phasediagram}).
At length scales shorter than $\xi_I$, the system retains the properties of a solid.
At intermediate scales, $\xi_I<L<\xi_{II}$, the system is smectic-like,
and consists of a regular stack of 1D channels of electron guiding centers, 
with short ranged translational order along the channels,
and quasi-long ranged order in the transverse direction.
Finally, at length scales longer than $\xi_{II}$, the system is nematic-like:
translational order is destroyed in all directions, but the system
preserves quasi long range orientational order. The latter is described
by the bond-angle field $\theta({\bf r})$,
which is defined as the orientation relative to some fixed reference axis of the bond
between two neighboring electron guiding centers. Standard analysis shows
that the fluctuations of $\theta({\bf r})$ are described by an effective Hamiltonian
of the form:
\begin{equation}
H_N = \frac{1}{2}\int d{\bf r}\,\big[ K_x(\partial_x\theta)^2 + K_y(\partial_y\theta)^2\big],
\end{equation}
and decay only algebraically with distance. Since on short length scales,
the AWC is only weakly disturbed and each electron is still surrounded on average by six neighbors,
he resulting quantum Hall ``nematic" state may be more accurately 
characterized as an {\em anisotropic hexatic}.

As the temperature is further raised, a disclination-unbinding
transition melts this nematic-like state into an isotropic metallic state,
in much the same way as described in Ref. \onlinecite{Ostlund1981}, 
with actual values of the nematic to isotropic melting temperature $T_{c2}$ of the order 
of those estimated in Ref. \onlinecite{Wexler2001}
(in this last reference, $T_{c2}\approx 200\,\mbox{mK}$ near half-filling of LL $N=2$).
Since the temperature ($\sim 25 \mbox{mK}$) at which the experiments of Ref. \onlinecite{Lilly1999}
were performed lies between $T_{c1}$ and $T_{c2}$, we see that our HF calculation
is consistent with the conjecture \cite{Kivelson1998,Fradkin1999}
according to which the state probed by these experiments is a nematic state.

{\em Conclusion} -- To summarize, in this paper we have 
examined the fluctuations and topological transitions of 
quantum Hall stripes near half-filling of intermediate Landau levels. 
Taking the stripe state to be an anisotropic Wigner crystal,
as suggested by Hartree-Fock and renormalization group calculations, we find that
the quantum Hall nematic conjectured in Refs. \cite{Kivelson1998,Fradkin1999} emerges 
in a natural way in the topological melting process, and is identified as an 
{\em anisotropic hexatic}. Our calculations are consistent with the idea of 
quantum Hall nematics, which we predict to be realized over
a significant region of the phase diagram near half filling of intermediate Landau levels, 
and give quantitative support to the qualitative interpretations \cite{Kivelson1998,Fradkin1999} 
of transport measurements \cite{Lilly1999} in terms of putative nematic states.

\medskip

\begin{acknowledgments}

AME acknowledges stimulating interaction with Hae Young Kee.
This work was supported by NSERC of Canada (AME, CBD, RC), 
and by FQRNT, the Swiss NSF and the NCCR Nanoscience (CBD).

\end{acknowledgments}

\end{document}